\documentclass[aip,jcp,reprint,showkeys]{revtex4-1}

\usepackage{graphicx,dcolumn,bm,xcolor,microtype,hyperref,multirow,amsmath,amssymb,amsfonts,physics,mathpazo,algorithmicx,algorithm,algpseudocode}
\usepackage[normalem]{ulem}
\usepackage[version=4]{mhchem}
\usepackage[compat=1.1.0]{tikz-feynman}

\algnewcommand\algorithmicassert{\texttt{assert}}
\algnewcommand\Assert[1]{\State \algorithmicassert(#1)}
\algrenewcommand{\algorithmiccomment}[1]{$\triangleright$ #1}

\definecolor{mypink}{rgb}{0.858, 0.188, 0.478}

\newcommand{\br}{\bm{r}}
\newcommand{\ree}{r_{12}}

\newcommand{\evGW}{evGW}	
\newcommand{\qsGW}{qsGW}	
\newcommand{\GOWO}{G$_0$W$_0$}

\newcommand{\GW}{GW}		
\newcommand{\GOWOSOSEX}{{\GOWO}+SOSEX}		
\newcommand{\GWSOSEX}{{\GW}+SOSEX}		
\newcommand{\GnWn}[1]{G$_{#1}$W$_{#1}$}		
\newcommand{\GOF}{G$_0$F2}			
\newcommand{\GF}{GF2}

\newcommand{\hH}{\Hat{H}}

\newcommand{\Ec}{E_\text{c}}
\newcommand{\EHF}{E_\text{HF}}

\newcommand{\EcRPA}{E_\text{c}^\text{RPA}}
\newcommand{\EcGM}{E_\text{c}^\text{GM}}
\newcommand{\EcGMGW}{E_\text{c}^\text{GM@GW}}
\newcommand{\EcGMGF}{E_\text{c}^\text{GM@GF2}}
\newcommand{\EcGMGWSOSEX}{E_\text{c}^\text{GM@GW+SOSEX}}
\newcommand{\EcMP}{E_c^\text{MP2}}

\newcommand{\Egap}{E_\text{gap}}
\newcommand{\IP}{\text{IP}}
\newcommand{\EA}{\text{EA}}

\newcommand{\nSat}[1]{N_{#1}^\text{sat}}
\newcommand{\eSat}[2]{\epsilon_{#1,#2}}
\newcommand{\e}[1]{\epsilon_{#1}}
\newcommand{\eHF}[1]{\epsilon^\text{HF}_{#1}}

\newcommand{\eGOWO}[1]{\epsilon^\text{\GOWO}_{#1}}

\newcommand{\eGnWn}[2]{\epsilon^\text{\GnWn{#2}}_{#1}}

\newcommand{\eGOF}[1]{\epsilon^\text{\GOF}_{#1}}
\newcommand{\de}[1]{\Delta\epsilon_{#1}}
\newcommand{\deHF}[1]{\Delta\epsilon^\text{HF}_{#1}}
\newcommand{\Om}[1]{\Omega_{#1}}
\newcommand{\eHOMO}{\epsilon_\text{HOMO}}
\newcommand{\eLUMO}{\epsilon_\text{LUMO}}

\renewcommand{\S}[1]{S_{#1}}
\newcommand{\ABSE}[1]{A^\text{BSE}_{#1}}
\newcommand{\BBSE}[1]{B^\text{BSE}_{#1}}
\newcommand{\ARPA}[1]{A^\text{RPA}_{#1}}
\newcommand{\BRPA}[1]{B^\text{RPA}_{#1}}
\newcommand{\dABSE}[1]{\delta A^\text{BSE}_{#1}}
\newcommand{\dBBSE}[1]{\delta B^\text{BSE}_{#1}}
\newcommand{\G}[1]{G_{#1}}
\newcommand{\Po}[1]{P_{#1}}
\newcommand{\W}[1]{W_{#1}}
\newcommand{\Wc}[1]{W^\text{c}_{#1}}
\newcommand{\vc}[1]{v_{#1}}
\newcommand{\SigX}[1]{\Sigma^\text{x}_{#1}}
\newcommand{\SigC}[1]{\Sigma^\text{c}_{#1}}
\newcommand{\SigGW}[1]{\Sigma^\text{\GW}_{#1}}
\newcommand{\SigGWSOSEX}[1]{\Sigma^\text{\GWSOSEX}_{#1}}
\newcommand{\SigGF}[1]{\Sigma^\text{\GF}_{#1}}
\newcommand{\Z}[1]{Z_{#1}}

\newcommand{\OmBSE}[1]{\Omega^\text{BSE}_{#1}}

\newcommand{\Oms}[1]{{}^{1}\Omega_{#1}}

\newcommand{\Omt}[1]{{}^{3}\Omega_{#1}}

\newcommand{\WJ}[3]{
\begin{pmatrix}
#1 & #2 & #3 \\
0  & 0  & 0  \\
\end{pmatrix} 
}
\newcommand{\ERI}[3]{\qty(#1 #2 #3)}
\newcommand{\sERI}[3]{\qty[#1 #2 #3]}

\newcommand{\bvc}{\bm{v}}

\newcommand{\bSigX}{\bm{\Sigma}^\text{x}}

\newcommand{\bSigGW}{\bm{\Sigma}^\text{\GW}}
\newcommand{\bSigGWSOSEX}{\bm{\Sigma}^\text{\GWSOSEX}}
\newcommand{\bSigGF}{\bm{\Sigma}^\text{\GF}}

\newcommand{\bDelta}{\bm{\Delta}}
\newcommand{\beHF}{\bm{\epsilon}^\text{HF}}

\newcommand{\beGnWn}[1]{\bm{\epsilon}^\text{\GnWn{#1}}}

\newcommand{\bde}{\bm{\Delta\epsilon}}

\newcommand{\bOm}{\bm{\Omega}}
\newcommand{\bA}{\bm{A}}
\newcommand{\bB}{\bm{B}}
\newcommand{\bX}{\bm{X}}
\newcommand{\bY}{\bm{Y}}
\newcommand{\bZ}{\bm{Z}}

\tikzfeynmanset{
every photon={red,thick},
every scalar={blue,thick}
}

\begin{document}	

\title{Green functions and self-consistency: insights from the spherium model}

\newcommand{\lcpq}{Laboratoire de Chimie et Physique Quantiques, Universit\'e de Toulouse, CNRS, UPS, France}
\newcommand{\lpt}{Laboratoire de Physique Th\'eorique, Universit\'e de Toulouse, CNRS, UPS, France}
\newcommand{\etsf}{European Theoretical Spectroscopy Facility (ETSF)}
\affiliation{\lcpq}
\affiliation{\lpt}
\affiliation{\etsf}

\author{Pierre-Fran{\c c}ois Loos}
\email[Corresponding author: ]{loos@irsamc.ups-tlse.fr}
\affiliation{\lcpq}
\author{Pina Romaniello}
\affiliation{\lpt}
\affiliation{\etsf}
\author{J.~A.~Berger}
\affiliation{\lcpq}
\affiliation{\etsf}

\begin{abstract}
We report an exhaustive study of the performance of different variants of Green function methods for the spherium model in which two electrons are confined to the surface of a sphere and interact via a genuine long-range Coulomb operator.
We show that the spherium model provides a unique paradigm to study electronic correlation effects from the weakly correlated regime to the strongly correlated regime, since the mathematics are simple while the physics is rich.
We compare perturbative GW, partially self-consistent GW and second-order Green function (GF2) methods for the computation of ionization potentials, electron affinities, energy gaps, correlation energies as well as singlet and triplet neutral excitations by solving the Bethe-Salpeter equation (BSE).
We discuss the problem of self-screening in GW and show that it can be partially solved with a second-order screened exchange correction (SOSEX).
We find that, in general, self-consistency deteriorates the results with respect to those obtained within perturbative approaches with a Hartree-Fock starting point.
Finally, we unveil an important problem of partial self-consistency in GW: in the weakly correlated regime, it can produce artificial discontinuities in the self-energy caused by satellite resonances with large weights.
\end{abstract}

\keywords{Green function; self-energy; GW approximation; self-consistent scheme}

\maketitle

\section{
\label{sec:intro}
Introduction}
\begin{figure}
	\includegraphics[width=0.7\linewidth]{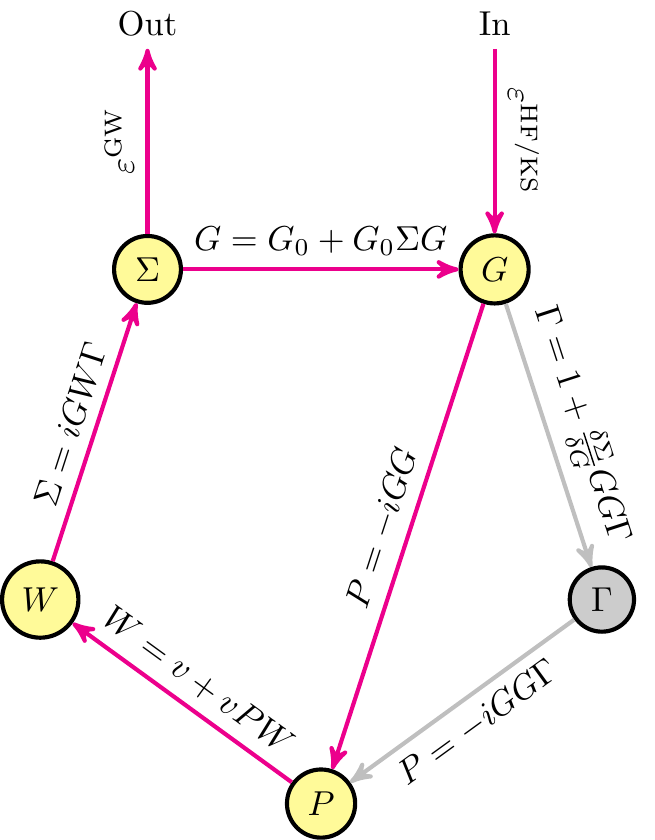}
	\caption{
	\label{fig:pentagon}
	Hedin's pentagon. \cite{Hedin_1965}
	The red path shows the self-consistent GW process which bypasses the computation of the vertex function $\Gamma$.}
\end{figure}

One-body Green function-based methods allow an explicit incorporation of electron correlation via a sequence of self-consistent steps, \cite{Hedin_1965} 
which connect the Green function $G$, the irreducible vertex function $\Gamma$, the irreducible polarizability $P$, the dynamically-screened Coulomb interaction $W$ 
and the self-energy $\Sigma$ through a set of five integro-differential equations known as Hedin's equations (see also Fig.~\ref{fig:pentagon}):
\begin{subequations}
\begin{align}
	\label{eq:G}
	& G(12) = G_\text{H}(12) + \int G_\text{H}(13) \Sigma(34) G(42) d(34),
	\\
	\label{eq:Gamma}
	& \Gamma(123) = \delta(12) \delta(13) 
	\notag 
	\\
	& \qquad \qquad		+ \int \fdv{\Sigma(12)}{G(45)} G(46) G(75) \Gamma(673) d(4567),
	\\
	\label{eq:P}
	& P(12) = - i \int G(13) \Gamma(324) G(41) d(34),
	\\
	\label{eq:W}
	& W(12) = v(12) + \int v(13) P(34) W(42) d(34),
	\\
	\label{eq:Sig}
	& \Sigma(12) = i \int G(13) W(14) \Gamma(324) d(34),
\end{align}
\end{subequations}
where $G_\text{H}$ is the one-body Hartree Green function, $v$ is the bare Coulomb interaction, $\delta(12)$ is Dirac's delta function \cite{NISTbook} and $(1)$ is a composite coordinate gathering spin, space and time variables $(\sigma_1,\br_1,t_1)$.
Important experimental properties such as ionization potentials, electron affinities as well as spectral functions, which are related to 
direct and inverse photo-emission, can be obtained directly from the one-electron Green function. \cite{Aryasetiawan_1998, Onida_2002, Reining_2017, Blase_2018, Danovich_2011, Ortiz_2013}
 
A particularly successful approximation to Hedin's equations in electronic-structure calculations is the so-called GW approximation \cite{Hedin_1965} 
which bypasses the calculation of the vertex corrections by setting \cite{Aryasetiawan_1998, Onida_2002, Reining_2017, Blase_2018} (see also Fig.~\ref{fig:pentagon}):
\begin{equation}
	\label{eq:GW}
	\Gamma(123) = \delta(12) \delta(13).
\end{equation}
Historically, GW methods have been mostly applied to solids. \cite{Aryasetiawan_1998, Onida_2002}
However, studies on atoms and molecules have been flourishing in the past ten years. \cite{Blase_2011, Faber_2011, Bruneval_2012, Bruneval_2013, Bruneval_2015, Bruneval_2016, Bruneval_2016a, Boulanger_2014, Blase_2016, Li_2017, Hung_2016, Hung_2017, vanSetten_2015, vanSetten_2018, Ou_2016, Ou_2018, Faber_2014}
Nowadays, efficient implementations of GW methods for localized basis sets are available in several softwares, such as \textsc{fiesta}, \cite{Blase_2011, Blase_2018} \textsc{molgw}, \cite{Bruneval_2016} \textsc{turbomole}, \cite{vanSetten_2013, Kaplan_2015, Kaplan_2016, Krause_2017} \textsc{fhi-aims} \cite{Caruso_2012, Caruso_2013, Caruso_2013a, Caruso_2013b} and others.
 
There exists many flavours of GW. 
The simplest and most popular variant is perturbative GW, or {\GOWO}, \cite{Hybertsen_1985a, Hybertsen_1986} which has been widely used in the literature to study solids, atoms and molecules. \cite{Bruneval_2012, Bruneval_2013, vanSetten_2015, vanSetten_2018}
Although {\GOWO} provides accurate results (at least for weakly/moderately correlated systems), it is strongly starting-point dependent due to its perturbative nature, and violates some important conservation laws. \cite{Onida_2002, Reining_2017}
Improvement may be obtained via partial or full self-consistency.

However, things are not that simple, as self-consistency and vertex corrections are known to cancel to some extent. \cite{Martin_2016}
Indeed, there is a long-standing debate about the importance of partial and full self-consistency in GW. \cite{Stan_2006, Stan_2009, Rostgaard_2010, Caruso_2012, Caruso_2013, Caruso_2013a, Caruso_2013b, Koval_2014, Wilhelm_2018}
In some situations, it has been found that self-consistency can worsen spectral properties compared to the simpler {\GOWO} method.
A famous example has been provided by the calculations performed on the uniform electron gas, \cite{Holm_1998, Holm_1999,Holm_2000, Garcia-Gonzalez_2001} 
a paradigm central to many areas of physics and chemistry. \cite{Loos_2016}
This was further evidenced in real extended systems by several authors. \cite{Schone_1998, Ku_2002, Kutepov_2016, Kutepov_2017}
However, other approximations may have caused such deterioration, e.g.~pseudo-potentials \cite{deGroot_1995} or finite-basis set effects. \cite{Friedrich_2006}
These studies have cast doubt on the importance of self-consistent schemes within GW, at least for solid-state calculations.
For finite systems such as atoms and molecules, the situation is less controversial, and partially or fully self-consistent GW methods have shown great promise. \cite{Ke_2011, Blase_2011, Faber_2011, Caruso_2012, Caruso_2013, Caruso_2013a, Caruso_2013b, Koval_2014, Hung_2016, Blase_2018, Jacquemin_2017}

Unfortunately, it is somewhat difficult to obtain reliable benchmark results for molecular systems as there are always inherent errors introduced by the one-electron basis set incompleteness, the pseudopotentials or additional numerical procedures such as Fourier transforms or the resolution of the identity approximation.
In that regard, exactly or very accurately solvable models have ongoing value, and are valuable both for illuminating the physics of more complicated systems and for testing theoretical approaches. \cite{Lani_2012, Berger_2014a, Berger_2014, Stan_2015, Tarantino_2017, Tarantino_2018}
Besides, they offer unparalleled mathematical simplicity, while retaining much of the key physics. \cite{Loos_2009a, Loos_2009c, Loos_2010, Loos_2010e}
One such model consists of two electrons, interacting through the long-range Coulomb potential but confined to the surface of a sphere, whose radius $R$ can be tuned to mimic weakly correlated systems ($R \ll 1$) or strongly correlated systems ($R \gg 1$). \cite{Seidl_2007, Loos_2009a} 
This paradigm possesses a number of interesting features, but the one of relevance here is that, for such system, it is possible to compute the exact or near-exact properties of the one-, two- and three-electron systems.
Additionally, one can obtain, like in the Hubbard model, most of the quantities of interest analytically, and the electronic interaction is, unlike the Hubbard model, genuinely long range.
Therefore, the ``two-electrons-on-a-sphere'' model --- dubbed ``spherium'' in the remaining of the paper --- can be seen as a unique theoretical laboratory to test the performances of the different GW variants.

The spherium model has already be considered by Schindlmayr \cite{Schindlmayr_2013} within the simple {\GOWO} method.
In particular, he reported analytical expressions for various quantities, such as the independent-particle Green function, 
the dynamically-screened Coulomb interaction and the self-energy.
He also studied the accuracy of {\GOWO} for the prediction of the HOMO-LUMO energy gap for various $R$ values, and provided a detailed analysis of the convergence behavior of the energy gap with respect to the size of the one-electron basis set.

Here, we propose to extend the analysis of Schindlmayr~\cite{Schindlmayr_2013} to unveil some interesting properties of self-consistent GW methods.
In particular, we compare {\GOWO}, partially self-consistent GW and second-order Green function (GF2) methods for a wide range of properties including ionization potentials, electron affinities, energy gaps, correlation energies as well as singlet and triplet neutral excitations by solving the Bethe-Salpeter equation (BSE).
We also study a perturbative and self-consistent version of a second-order screened exchange correction (SOSEX) to the GW self-energy, labeled as GW+SOSEX.
We focus here on self-consistent schemes that are widely used and available, for example, through the software packages mentioned above.
For this reason fully self-consistent GW is beyond the scope of this work.

The paper is organised as follows. 
In Sec.~\ref{sec:paradigm}, we briefly review the GW equations for spherium. 
Section~\ref{sec:green} provides details about our perturbative and self-consistent GW implementations, and give the expression of the self-energy for GF2 and GW+SOSEX. 
We also report the expression of the BSE singlet/triplet excitations and various energy functionals. 
Results are reported and discussed in Sec.~\ref{sec:results}. 
Finally, we draw our conclusions in Sec.~\ref{sec:conclusion}. 
Atomic units are used throughout.

\section{
\label{sec:paradigm}
Two electrons on a sphere}
In this section we briefly review the GW equations for the spherium model, which consists of two opposite-spin electrons restricted to remain on the surface of a sphere of radius $R$.
\cite{Seidl_2007, Loos_2009, Loos_2009a, Loos_2009c, Loos_2010, Loos_2010d, Loos_2010e, Loos_2011b, Loos_2012, Gill_2012, Loos_2012c, Loos_2015b, Loos_2017a}
The Hamiltonian of the system is simply
\begin{equation}
	\hH = -\frac{\nabla_1^2+\nabla_2^2}{2} + \frac{1}{\ree},
\end{equation}
where 
\begin{equation}
	\nabla_i^2 = \frac{1}{R^2} \qty( \pdv[2]{}{\theta_i} + \cot \theta_i \pdv{}{\theta_i} + \frac{1}{\sin^2 \theta_i} \pdv[2]{}{\phi_i} )
\end{equation}
is the angular part of the Laplace operator for electron $i$ and $\ree = \abs{\br_1 - \br_2}$ is the distance between the two electrons i.e., the electrons interact Coulombically \emph{through} the sphere.
Note that we eschew the introduction of a positively-charged background, which is equivalent to a trivial energy shift.

The Hartree-Fock (HF) orbitals of spherium are the normalized spherical harmonics $Y_{\ell m}(\theta,\phi)/R$, where $ 0 \le \ell \le L$, $-\ell \le m \le +\ell$, $L$ is the maximum angular momentum of the one-electron basis set, and $(\theta,\phi)$ are the polar and azimuthal angles respectively.
We will use this convenient, orthogonal and complete basis set to represent the various quantities associated with GW methods.
Moreover, because we focus our attention on the totally-symmetric singlet ground state, all the quantities of interest are independent of $m$.
Therefore, from hereon, we drop the $m$ dependence. \cite{Loos_2009a}
A crucial point here is that, as shown by Schindlmayr, \cite{Schindlmayr_2013} all the quantities reported in this section have a diagonal representation, i.e., only their diagonal elements are non-zero.
As we shall see below, this yields important simplifications in the GW equations and their implementation.

The HF orbital energies are given by 
\begin{equation}
\label{eq:eHF}
	\eHF{\ell} = \frac{\ell(\ell+1)}{2R^2} + \frac{2}{R} + \SigX{\ell},
\end{equation}
where the exchange part of the self-energy is
\begin{equation}
\label{eq:SigX}
	\SigX{\ell} = - \frac{1}{(2\ell+1) R}.
\end{equation}
Within the single-determinant approximation, in its singlet ground state, the lowest $s$-type spherical harmonic $Y_{00}(\theta,\phi)$ is doubly occupied and the electron density is uniform over the sphere: \cite{Loos_2011b}
\begin{equation}
	\rho = 2 \abs{Y_{00}(\theta,\phi)/R}^2 = \frac{1}{2\pi R^2}.
\end{equation}
All the GW calculations reported in this study have been performed with a HF starting point.

As derived by Schindlmayr, \cite{Schindlmayr_2013} the independent-particle Green function is
\begin{equation}
	\label{eq:GHF}
	\G{\ell}(\omega) = \frac{\delta_{0\ell}}{\omega - \e{\ell} - i \eta} + \frac{1 - \delta_{0\ell}}{\omega - \e{\ell} + i \eta},
\end{equation}
(where $\eta$ is a positive infinitesimal and $\delta_{\ell_1 \ell_2}$ is the Kronecker delta \cite{NISTbook}) and the polarizability function reads
\begin{equation}
	\label{eq:P}
	\Po{\ell}(\omega) = \frac{1 - \delta_{0\ell}}{2\pi R^2} \qty(\frac{1}{\omega - \de{\ell} + i \eta} - \frac{1}{\omega + \de{\ell} - i \eta} ),
\end{equation}
with
\begin{equation}
	\label{eq:de}
	\de{\ell} = \e{\ell} - \e{0}.
\end{equation}
Defining 
\begin{equation}
	\label{eq:W}
	\W{\ell}(\omega) = \vc{\ell} + \Wc{\ell}(\omega),
\end{equation}
with
\begin{equation}
	\label{eq:vc}
	\vc{\ell} = \frac{4\pi}{2\ell+1} R,
\end{equation}
the correlation part of the dynamically-screened Coulomb interaction is
\begin{multline}
	\label{eq:Wc}
	\Wc{\ell}(\omega) = \frac{(1 - \delta_{0\ell}) \vc{\ell}^2}{2\pi R} \frac{\de{\ell}}{\Om{\ell}} 
	\\
	\times \qty(\frac{1}{\omega - \Om{\ell} + i \eta} - \frac{1}{\omega + \Om{\ell} - i \eta} ),
\end{multline}
where
\begin{equation}
\label{eq:Omega}
	\Om{\ell} = \sqrt{\de{\ell}(\de{\ell} - 4\SigX{\ell})}
\end{equation}
are the (singlet) random phase approximation (RPA) excitation energies.
Defining, respectively, the bare and screened two-electron integrals as
\begin{subequations}
\begin{align}
	\label{eq:ERI}
	\ERI{\ell_1}{\ell_2}{\ell}  & = \frac{1}{R} \sqrt{\frac{\vc{\ell_2}}{\vc{\ell_1}}} \WJ{\ell_1}{\ell_2}{\ell},
	\\
	\label{eq:sERI}
	\sERI{\ell_1}{\ell_2}{\ell} & = \sqrt{\frac{\de{\ell_2}}{\Om{\ell_2}}} \ERI{\ell_1}{\ell_2}{\ell},
\end{align}
\end{subequations}
this yields 
\begin{equation}
\begin{split}
	\label{eq:SigC_GW}
	\SigGW{\ell}(\omega)
	& = \frac{2(1 - \delta_{0\ell}) \sERI{0}{\ell}{\ell}^2}{\omega - \e{0} + \Om{\ell} - i \eta}
	\\
	& + \sum_{\ell_1,\ell_2=1}^{L} \frac{2\sERI{\ell_1}{\ell_2}{\ell}^2}{\omega - \e{\ell_1} - \Om{\ell_2} + i \eta}
\end{split}
\end{equation}
for the correlation part of the GW self-energy, where 
\begin{multline}
	\label{eq:W3J}
	\WJ{\ell_1}{\ell_2}{\ell}^2 = 
	\sqrt{\frac{4\pi}{(2\ell_1 +1)(2\ell_2 + 1)(2\ell + 1)}} 
	\\
	\times \int_0^{2\pi} \int_0^\pi Y_{\ell_1 0}(\theta,\phi) Y_{\ell_2 0}(\theta,\phi) Y_{\ell 0}(\theta,\phi) \sin \theta d\theta d\phi
\end{multline}
defines the Wigner 3j symbol. \cite{NISTbook}
More details about the derivation of all these quantities can be found in Ref.~\onlinecite{Schindlmayr_2013}.

\begin{figure}
	\begin{algorithmic}[1]
		\Procedure{Self-consistent {\GW}}{}
			\State Precompute $\bSigX$, $\bvc$ and Wigner 3j symbols via Eqs.~\eqref{eq:SigX}, \eqref{eq:W3J} and \eqref{eq:vc}, respectively
			\State Set $\beGnWn{-1} = \beHF$  and $n = 0$
			\While{$\max{\abs{\bDelta}} < \tau$}
				\State Form $\bde$ (Eq.~\eqref{eq:de}) and $\bOm$ (Eq.~\eqref{eq:Omega}) using $\beGnWn{n-1}$
				\State Form $\bSigGW(\omega)$ following Eq.~\eqref{eq:SigC_GW}
				\For{$\ell=1,\ldots,L$}
					\State Solve $\omega = \eHF{\ell} + \Re[\SigGW{\ell}(\omega)]$ to obtain $\eGnWn{\ell}{n}$
				\EndFor
				\State $\bDelta = \beGnWn{n} - \beGnWn{n-1}$
				\State $n \leftarrow n + 1$
			\EndWhile 
			\If{BSE}
				\State Compute BSE excitations energies (see Table \ref{tab:Ex})
			\EndIf
		\EndProcedure
	\end{algorithmic}
	\caption{
		\label{fig:algo_GW}
		Pseudo-code for self-consistent {\GW} calculations.
		$\tau$ is a user-defined threshold.
	}
\end{figure}

\section{
\label{sec:green}
Green function methods}
\subsection{\GOWO}

In {\GOWO}, one only updates once the orbital energies, which are obtained by solving a linearized (static) version of the quasiparticle equation \cite{Hybertsen_1985a, Hybertsen_1986}
\begin{equation}
\label{eq:eG0W0}
	\eGOWO{\ell} = \eHF{\ell} + \Z{\ell}(\eHF{\ell})\,\Re[\SigGW{\ell}(\eHF{\ell})],
\end{equation}
where $\eHF{\ell}$ and $\SigGW{\ell}$ are given by Eqs.~\eqref{eq:eHF} and \eqref{eq:SigC_GW} respectively, and the renormalization factor
\begin{equation}
	\label{eq:Z}
	\Z{\ell}(\omega) = \qty[ 1 - \pdv{\Re[\SigGW{\ell}(\omega)]}{\omega} ]^{-1}
\end{equation}
specifies the weight of the quasiparticle energy in the spectral function
\begin{equation}
	\S{\ell}(\omega) = \pi^{-1} \abs{\Im[\G{\ell}(\omega)]}.
\end{equation}
\subsection{Self-consistent GW}
As mentioned in Sec.~\ref{sec:intro}, the major drawback of {\GOWO} is its starting point dependency.
One way of getting rid of this shortcoming is to iterate the GW quantities until self-consistency has been reached.
The important point here is that, thanks to the diagonal nature of all the GW quantities (see Sec.~\ref{sec:paradigm}), their expressions are valid (within the quasiparticle approximation), not only at the zeroth iteration, but at any stage of the self-consistent iterative scheme.

There exists two main types of partially self-consistent GW methods: i) in \textit{``eigenvalue-only quasiparticle''} GW ({\evGW}), \cite{Hybertsen_1986, Shishkin_2007, Blase_2011, Faber_2011} the quasiparticle energies are updated at each iteration; ii) in \textit{``quasiparticle self-consistent''} GW ({\qsGW}), \cite{Faleev_2004, vanSchilfgaarde_2006, Kotani_2007, Ke_2011} one updates both the quasiparticle energies and the corresponding orbitals.
Note that a starting point dependency remains in {\evGW} as the orbitals are not self-consistency optimized in that case.
However, in the present model, thanks to the diagonal nature of the various intermediates, the orbitals do not mix from one iteration to another, and the only ``updatable'' quantities are the quasiparticle energies.
Consequently, the partially self-consistent GW schemes {\evGW} and {\qsGW} are strictly equivalent.
Hence, we will not be making any distinction between them and label them as {\GW} in the following.

A pseudo-code of the self-consistent GW algorithm is reported in Fig.~\ref{fig:algo_GW}.
In the present implementation, at the $n$th iteration, the {\GW} quasiparticle orbital energies $\eGnWn{\ell}{n}$ are determined by solving the non-linear, frequency-dependent quasiparticle equation
\begin{equation}
\label{eq:QP-eq}
	\omega = \eHF{\ell} + \Re[\SigGW{\ell}(\omega)].
\end{equation}
Note that $\bSigGW$ is built with the orbital energy differences $\bde$ and RPA excitation energies $\bOm$ computed with the orbital energies from the previous iteration, i.e.~$\beGnWn{n-1}$.
The self-consistent process is carried on until the convergence criterion
\begin{equation}
	\max{\abs{\beGnWn{n} - \beGnWn{n-1}}} < \tau
\end{equation}
is met (where $\tau$ is a user-defined threshold).

The various solutions of the quasiparticle equation \eqref{eq:QP-eq}, $\omega_{\ell,s}$, have different meanings.
For each $\ell$ value, in addition to the principal quasiparticle energy $\eSat{\ell}{0} \equiv \e{\ell}$, there is a finite number of satellites resonances $\nSat{\ell}$ at frequencies $\eSat{\ell}{s}$ ($s>0$) stemming from the poles of the self-energy. 
One can show that the two following sum rules \cite{vonBarth_1996} are fulfilled:
\begin{align}
		\label{eq:sumrules}
		\sum_{s=0}^{\nSat{\ell}} Z_{\ell}(\eSat{\ell}{s}) & = 1,
		& 
		\sum_{s=0}^{\nSat{\ell}} Z_{\ell}(\eSat{\ell}{s}) \eSat{\ell}{s} & = \eHF{\ell},
\end{align}
where $Z_{\ell}(\omega)$ is given by Eq.~\eqref{eq:Z}. 

In a weakly or moderately correlated regime, one can clearly distinguish dominant quasiparticle peaks from satellites, whereas this scenario can break down in the strongly correlated regime. 
However, as we shall see below, this is not always the case.
In the present quasiparticle GW scheme, one only updates the quasiparticle energies, and the satellite resonances are discarded. 
Hence, the quasiparticle weights are reset to one at each iteration.

\begin{figure}
\begin{align*}
\bSigGW & = 
\feynmandiagram[baseline=(f3.base),small,horizontal=f1 to f2] {
f1 -- [fermion, quarter left] f2, 
f2 -- [fermion, quarter left] f1,
f4 -- [anti fermion] f3,
f1 -- [scalar] f4,
f3 -- [photon] f2
};
=
\feynmandiagram[baseline=(f3.base),small,horizontal=f1 to f2] {
f1 -- [fermion, quarter left] f2, 
f2 -- [fermion, quarter left] f1,
f4 -- [anti fermion] f3,
f1 -- [scalar] f4,
f3 -- [scalar] f2
};
+
\feynmandiagram[baseline=(f7.base),small,horizontal=f1 to f2] {
f1 -- [draw=none] f2 -- [fermion, bend left] f3, 
f3 -- [fermion, bend left] f2,
f4 -- [fermion, bend left] f5 -- [draw=none] f6, 
f5 -- [fermion, bend left] f4,
f7 --  f8 -- [anti fermion] f9,
f2 -- [scalar] f5,
f3 -- [scalar] f6,
f4 -- [scalar] f7,
f6 -- [scalar] f9
};
+\ldots
\\
\bSigGF & = 
\feynmandiagram [horizontal=f1 to f2,small] {
f1 -- [fermion, quarter left] f2, 
f2 -- [fermion, quarter left] f1,
f4 -- [anti fermion] f3,
f1 -- [scalar] f4,
f3 -- [scalar] f2
};
+
\feynmandiagram [horizontal=f1 to f2,small] {
f1 -- [anti fermion] f2 -- [anti fermion] f3 -- [anti fermion] f4, 
f1 -- [scalar, half left] f3,
f2 -- [scalar, half left] f4
};
\\
\bSigGWSOSEX & =
\feynmandiagram [horizontal=f1 to f2,small] {
f1 -- [fermion, quarter left] f2, 
f2 -- [fermion, quarter left] f1,
f4 -- [anti fermion] f3,
f1 -- [scalar] f4,
f3 -- [photon] f2
};
+
\feynmandiagram [horizontal=f1 to f2,small] {
f1 -- [anti fermion] f2 -- [anti fermion] f3 -- [anti fermion] f4, 
f1 -- [scalar, half left] f3,
f2 -- [photon, half left] f4
};
\end{align*}
\caption{
	\label{fig:diagrams} 
	Diagrammatic representation of the correlation part of the GW, GF2 and GW+SOSEX self-energies.
	Arrowed solid black lines, dashed blue lines and wiggly red lines indicate the one-body Green function $G$, the bare Coulomb interaction $v$, and dynamically-screened Coulomb interaction $W$, respectively
	In perturbative or self-consistent calculations, the propagator $G$ is bared or dressed, respectively.
	}
\end{figure}
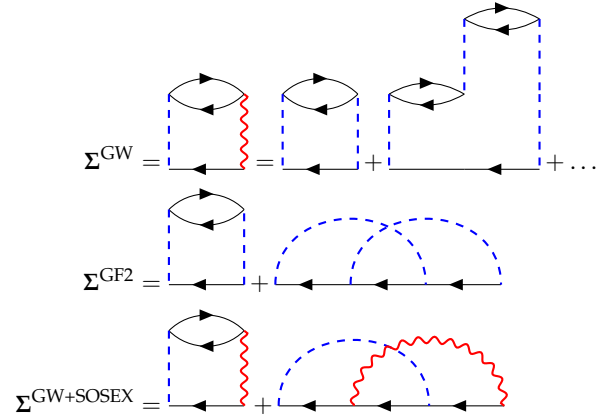

\subsection{GF2}
Diagrammatically, the difference between GW and GF2 is simple to explain: while GW takes into account all the direct ring diagrams, GF2 only includes the two (direct and exchange) second-order diagrams. \cite{Ortiz_2013, Hirata_2015, Hirata_2017}
Therefore, GF2 does not take into account the screening of the Coulomb interaction. 
This is illustrated in Fig.~\ref{fig:diagrams} in terms of Feynman diagrams. 
Note that GF2 is also known as the second Born approximation. \cite{Stefanucci_2013}
Like in GW, the correlation part of the GF2 self-energy has a diagonal representation:
\begin{multline}
\label{eq:SigC_GF2}
\begin{split}
	\SigGF{\ell}(\omega) 
	& = \frac{(1 - \delta_{0\ell}) \ERI{0}{\ell}{\ell}^2}{\omega - \e{0} + \de{\ell} - i \eta}
	\\
	& + \sum_{\ell_1,\ell_2=1}^{L} 
	\frac{2\ERI{\ell_1}{\ell_2}{\ell}^2 - \ERI{\ell_1}{\ell_2}{\ell}\ERI{\ell_2}{\ell_1}{\ell}}{\omega - \e{\ell_1} - \de{\ell_2} + i \eta}.
\end{split}
\end{multline}

Similarly to {\GOWO}, we consider a one-shot, perturbative GF2 procedure.
For sake of consistency, we will label these calculations as {\GOF}. 
The {\GOF} quasiparticle orbital energies $\eGOF{\ell}$ are given by Eq.~\eqref{eq:eG0W0} where one replaces $\SigGW{\ell}$ by $\SigGF{\ell}$, with a similar substitution for the renormalization factor reported in Eq.~\eqref{eq:Z}.
We also consider a self-consistent version.
The self-consistent procedure for GF2 is similar to the self-consistent GW scheme detailed in Fig.~\ref{fig:algo_GW}, except that one substitutes the GW self-energy \eqref{eq:SigC_GW} by its GF2 counterpart given by Eq.~\eqref{eq:SigC_GF2}.
From hereon, we will label these self-consistent calculations as {\GF}.

\subsection{GW+SOSEX}
To combine the best of both worlds, we propose to study a combination of GW and GF2, which is equivalent to the GW+SOSEX method recently introduced by Ren et al. \cite{Ren_2015}
GW+SOSEX is a well-defined diagrammatic method which adds, like in the SOSEX version of RPA, a subset of higher-order exchange-type diagrams to the formally infinite number of direct ring diagrams from GW (see Fig.~\ref{fig:diagrams}). \cite{Ren_2012}
Unlike Ren et al., \cite{Ren_2015} we test both the one-shot version, labeled {\GOWOSOSEX}, as well as its self-consistent version {\GWSOSEX}.
The correlation part of the GW+SOSEX self-energy is given by
\begin{equation}
\begin{split}
\label{eq:SigC_GWSOSEX}
	\SigGWSOSEX{\ell}(\omega) 
	& = \frac{(1 - \delta_{0\ell})\sERI{0}{\ell}{\ell}^2}{\omega - \e{0} + \Om{\ell} - i \eta}
	\\
	& + \sum_{\ell_1,\ell_2=1}^{L} 
	\frac{2\sERI{\ell_1}{\ell_2}{\ell}^2-\sERI{\ell_1}{\ell_2}{\ell}\sERI{\ell_2}{\ell_1}{\ell}}{\omega - \e{\ell_1} - \Om{\ell_2} + i \eta},
\end{split}
\end{equation}
which corresponds to the GF2 expression \eqref{eq:SigC_GF2} where one has substituted the bare two-electron integrals (Eq.~\eqref{eq:ERI}) by their screened version (Eq.~\eqref{eq:sERI}) stemming from the GW self-energy expression \eqref{eq:SigC_GW}.
As we shall see later on, the main advantage of this hybrid method is to partially remove self-screening which hampers the accuracy of the GW method, in particular for few-electron systems. \cite{Romaniello_2012, Aryasetiawan_2012, Wetherell_2018}
Again, the implementation of GW+SOSEX follows closely the algorithm detailed in Fig.~\ref{fig:algo_GW}, except that one replaces the GW self-energy \eqref{eq:SigC_GW} by its SOSEX-corrected version given by Eq.~\eqref{eq:SigC_GWSOSEX}.

\begin{table*}
\caption{
\label{tab:Ex}
Expression of the $\ell$th singlet excitation energy $\Oms{\ell}$ and $\ell$th triplet excitation energy $\Omt{\ell}$ for various methods.
The lowest excitation corresponds to $\ell = 1$.}
\begin{ruledtabular}
\begin{tabular}{lcc}
Method    	&	Singlet excitation energies $\Oms{\ell}$	&	Triplet excitation energies $\Omt{\ell}$	\\
\hline
CIS		&	$\deHF{\ell} - 2\SigX{\ell} + \SigX{0}$
		&	$\deHF{\ell} + \SigX{0}$
		\\
TDHF	&	$\sqrt{\qty(\deHF{\ell} - \SigX{\ell} + \SigX{0})\qty(\deHF{\ell} - 3\SigX{\ell} + \SigX{0})}$
		&	$\sqrt{\qty(\deHF{\ell} - \SigX{\ell} + \SigX{0})\qty(\deHF{\ell} + \SigX{\ell} + \SigX{0})}$
		\\
RPA		&	$\sqrt{\de{\ell}\qty(\de{\ell} - 4\SigX{\ell})}$
		&	$\de{\ell}$
		\\
BSE		&	$\sqrt{\qty(\de{\ell} - \SigX{\ell} + \SigX{0} - \frac{4 (\SigX{\ell})^2}
			{\de{\ell} - 4\SigX{\ell}})\qty(\de{\ell} - 3 \SigX{\ell} + \SigX{0} + \frac{4 (\SigX{\ell})^2}{\de{\ell} - 4\SigX{\ell}})}$
		&	$\sqrt{\qty(\de{\ell} - \SigX{\ell} + \SigX{0} - \frac{4 (\SigX{\ell})^2}{\de{\ell}})\qty(\de{\ell} + \SigX{\ell} + \SigX{0} + \frac{4 (\SigX{\ell})^2}{\de{\ell}})}$
				\\
\end{tabular}
\end{ruledtabular}
\end{table*}

\subsection{Bethe-Salpeter equation}
From the first-order variation of $G$ with respect to a general non-local external potential, one can get the neutral excitations of the system. 
This corresponds to solving the Bethe-Salpeter equation (BSE). \cite{Strinati_1988} 
Here, we use the BSE within the GW approximation (BSE@GW) \cite{Leng_2016, Blase_2018} to study the singlet and triplet neutral excitations of spherium. 
Note that the BSE calculations are performed as a post-GW step (see Fig.~\ref{fig:algo_GW}). 
We compare BSE with two common quantum chemistry methods: configuration interaction singles (CIS) and time-dependent HF (TDHF). 
We refer the interested readers to the review of Dreuw and Head-Gordon for more details about these conventional methods. \cite{Dreuw_2015}

Thanks to the unique feature of the present model, the linear response eigenvalue problem has a diagonal structure.
It implies that the eigenvalues and eigenvectors can be trivially obtained, and the singlet and triplet excitation energies can be easily written in closed form for all the methods mentioned above. 
Their expressions are gathered in Table \ref{tab:Ex}.

To compute the BSE excitation energies for the singlet and triplet manifolds, one must solve the following linear response problem \cite{Dreuw_2005}
\begin{equation}
	\begin{pmatrix}
		\bA	&	\bB	\\
		\bB	&	\bA	\\
	\end{pmatrix}
	\begin{pmatrix}
		\bX	\\
		\bY	\\
	\end{pmatrix}
	=
	\bOm 
	\begin{pmatrix}
		\bm{1}	&	0	\\
		0	&	\bm{-1}	\\
	\end{pmatrix}
	\begin{pmatrix}
		\bX	\\
		\bY	\\
	\end{pmatrix},
\end{equation}
which is usually transformed (when the orbitals do not exhibit triplet instabilities \cite{Dreuw_2005}) into an eigenvalue problem of smaller dimension
\begin{equation}
\label{eq:small-LR}
	(\bA - \bB)^{1/2} (\bA + \bB) (\bA - \bB)^{1/2} \bZ = \bOm^2 \bZ,
\end{equation}
where the excitation amplitudes are
\begin{equation}
	\bX + \bY = \bOm^{-1/2} (\bA - \bB)^{1/2} \bZ.
\end{equation}
The only difference between CIS, TDHF, RPA and BSE lies in the specific expression of the matrix elements of $\bA$ and $\bB$. 
As mentioned above, in the present case, the matrices $\bA$ and $\bB$ have a diagonal structure, and the BSE diagonal elements are given by
\begin{align}
	\ABSE{\ell} & = \ARPA{\ell} + \dABSE{\ell},
	&
	\BBSE{\ell} & = \BRPA{\ell} + \dBBSE{\ell},
\end{align}
where the RPA part is
\begin{align}
	\label{eq:LR_RPA}
	\ARPA{\ell} & = \de{\ell} - 2 (1 - \delta_{\sigma \sigma^{\prime}}) \SigX{\ell},
	&
	\BRPA{\ell} & = - 2 (1 - \delta_{\sigma \sigma^{\prime}}) \SigX{\ell},
\end{align}
and the BSE correction reads 
\begin{align}
	\label{eq:LR_BSE}
	\dABSE{\ell} & = \SigX{0},
	& 
	\dBBSE{\ell} & = \SigX{\ell} - \frac{4 (\SigX{\ell})^2}{\ARPA{\ell}+\BRPA{\ell}},
\end{align}
with
\begin{equation}
	\delta_{\sigma \sigma^{\prime}} = 
	\begin{cases}
		0,	&	\sigma \neq \sigma^{\prime} \text{ (singlet manifold)},
		\\
		1,	&	\sigma = \sigma^{\prime} \text{ (triplet manifold)}.
	\end{cases}
\end{equation}
Therefore, substituting \eqref{eq:LR_RPA} into \eqref{eq:small-LR} yields the BSE excitation energy:
\begin{equation}
	\OmBSE{\ell} = \sqrt{\qty(\ABSE{\ell} - \BBSE{\ell})\qty(\ABSE{\ell} + \BBSE{\ell})}.
\end{equation}
Their explicit expression for the singlet and triplet manifold are provided in Table \ref{tab:Ex}.
The CIS, TDHF and RPA excitations energies can be obtained via the same derivation and their expressions are also reported in Table \ref{tab:Ex}.

\subsection{Correlation energy}
The correlation energy is defined as
\begin{equation}
	\Ec = E - \EHF,
\end{equation}
where $E$ is a total energy estimate provided by a given correlated method and $\EHF = 1/R$ is the (restricted) HF energy of the singlet ground-state of spherium. \cite{Loos_2009a}

We followed two distinct routes to estimate the correlation energy within GW.
First, we estimated the correlation energy within the RPA: \cite{Casida_1995, Dahlen_2006, Furche_2008, Bruneval_2016}
\begin{equation}
	\label{eq:Ec-RPA}
	\EcRPA = -\sum_{\ell=1}^L \qty(\ARPA{\ell} - \Om{\ell}),
\end{equation}
where the $\ARPA{\ell}$'s and $\Om{\ell}$'s are given by Eqs.~\eqref{eq:LR_RPA} and \eqref{eq:Omega}, respectively.
We note that Eq.~\eqref{eq:Ec-RPA} can be obtained from the variational Klein functional~\cite{Klein_1961} if the GW approximation is used for the Luttinger-Ward (or $\Phi$) functional. \cite{Luttinger_1960}

The second route we followed was to calculate the functional \cite{Holm_1998, Caruso_2013b}
\begin{equation}
	\label{eq:GM}
	\EcGM = \frac{-i}{2}\sum_{\ell=0}^{\infty}\int \frac{d\omega}{2\pi} \SigC{\ell}(\omega) \G{\ell}(\omega) e^{i\omega\eta}.
\end{equation}
This equation is equivalent to the correlation part of the Galitskii-Migdal (GM) functional~\cite{Galitskii_1958} for the total energy if the self-energy and the Green function are connected through the Dyson equation. For this reasons we will refer to the above expression as the GM functional for the correlation energy.
In our case, the frequency integration in Eq.~(\ref{eq:GM}) can be performed analytically. 
We obtain
\begin{subequations}
\begin{align}
	\label{eq:Ec-GMGW}
	\EcGMGW & = - 2 \sum_{\ell=1}^{L} \frac{\sERI{0}{\ell}{\ell}^2+\sERI{\ell}{\ell}{0}^2}{\de{\ell} + \Om{\ell}},
	\\
	\label{eq:Ec-GMGF2}
	\EcGMGF & = - \sum_{\ell=1}^{L} \frac{\ERI{0}{\ell}{\ell}^2+\ERI{\ell}{\ell}{0}^2}{\de{\ell} + \de{\ell}},		
	\\
	\label{eq:Ec-GMGWSOSEX}
	\EcGMGWSOSEX & = - \sum_{\ell=1}^{L} \frac{\sERI{0}{\ell}{\ell}^2+\sERI{\ell}{\ell}{0}^2}{\de{\ell} + \Om{\ell}}.
\end{align}
\end{subequations}
We note that $\EcGMGWSOSEX = \frac12\EcGMGW$. 
As discussed in the next section this is related to the self-screening problem of GW.
It is well known that, because we only consider partially self-consistent GW schemes, the correlation energy estimates provided by the RPA and GM functionals will differ, \cite{Stan_2006} while, at full self-consistency, they would be identical. \cite{Caruso_2013, Caruso_2013a, Caruso_2013b}

For comparison purposes, we have also computed the second-order M{\o}ller-Plesset (MP2) correlation energy, \cite{SzaboBook} which reads
\begin{equation}
	\label{eq:Ec-MP2}
	\EcMP = - \sum_{\ell=1}^{L} \frac{\ERI{0}{\ell}{\ell}^2}{\de{\ell} + \de{\ell}}.	
\end{equation}
The similarity between the MP2 expression \eqref{eq:Ec-MP2} and the GM expressions reported in Eqs.~\eqref{eq:Ec-GMGW}, \eqref{eq:Ec-GMGF2} and \eqref{eq:Ec-GMGWSOSEX} is striking.

\section{
\label{sec:results}
Results}
\subsection{Computational details}
In practice, one only requires the energy of the main quasiparticle peaks at each iteration (i.e.~the satellites can be discarded).
For each $\ell$ value, the quasiparticle energy is found by solving the quasiparticle equation (see, for example, Eq.~\eqref{eq:QP-eq}) using Newton's method (as implemented in Mathematica 11) starting from the result of the previous iteration.
In order to avoid finite-size basis set effects, the maximum angular momentum of the basis set has been set to $L = 50$, which ensures converged results with respect to the basis set size up to $R = 10$, the largest radius considered here.
However, we only update the eigenvalues for $0 \le \ell \le 10$ which corresponds to the HOMO ($\ell = 0$), the LUMO ($\ell = 1$) and the next nine unoccupied orbitals.
As mentioned earlier, all the calculations have been performed with a (restricted) HF starting point. \cite{Loos_2009a}
For the self-consistent GW calculations, the convergence threshold has been set to $\tau = 10^{-5}$.
In case of convergence issues, instead of the usual linear mixing performed in standard implementations, \cite{Caruso_2013a, Kaplan_2016} we have found that the DIIS extrapolation technique originally proposed by Pulay \cite{Pulay_1980, Pulay_1982} is more robust and rather efficient.

The quantities labeled as ``exact'' have been obtained from near-exact calculations computed with the full configuration interaction (FCI) method. \cite{Loos_2009a}
In particular, we have computed the exact energies of the one-, two- and three-electron systems for various $R$ values, as well as the singlet and triplet excitation energies of the two-electron system.
For some well-defined values of $R$ (such as $R = \sqrt{3}/2$ or $\sqrt{7}$), exact wave functions and energies are available in the case of the two-electron system. \cite{Loos_2009c, Loos_2012}
Finally, we note that we have verified that the sum rules in Eq.~(\ref{eq:sumrules}) are satisfied in our calculations.

\begin{figure}
	\includegraphics[width=\linewidth]{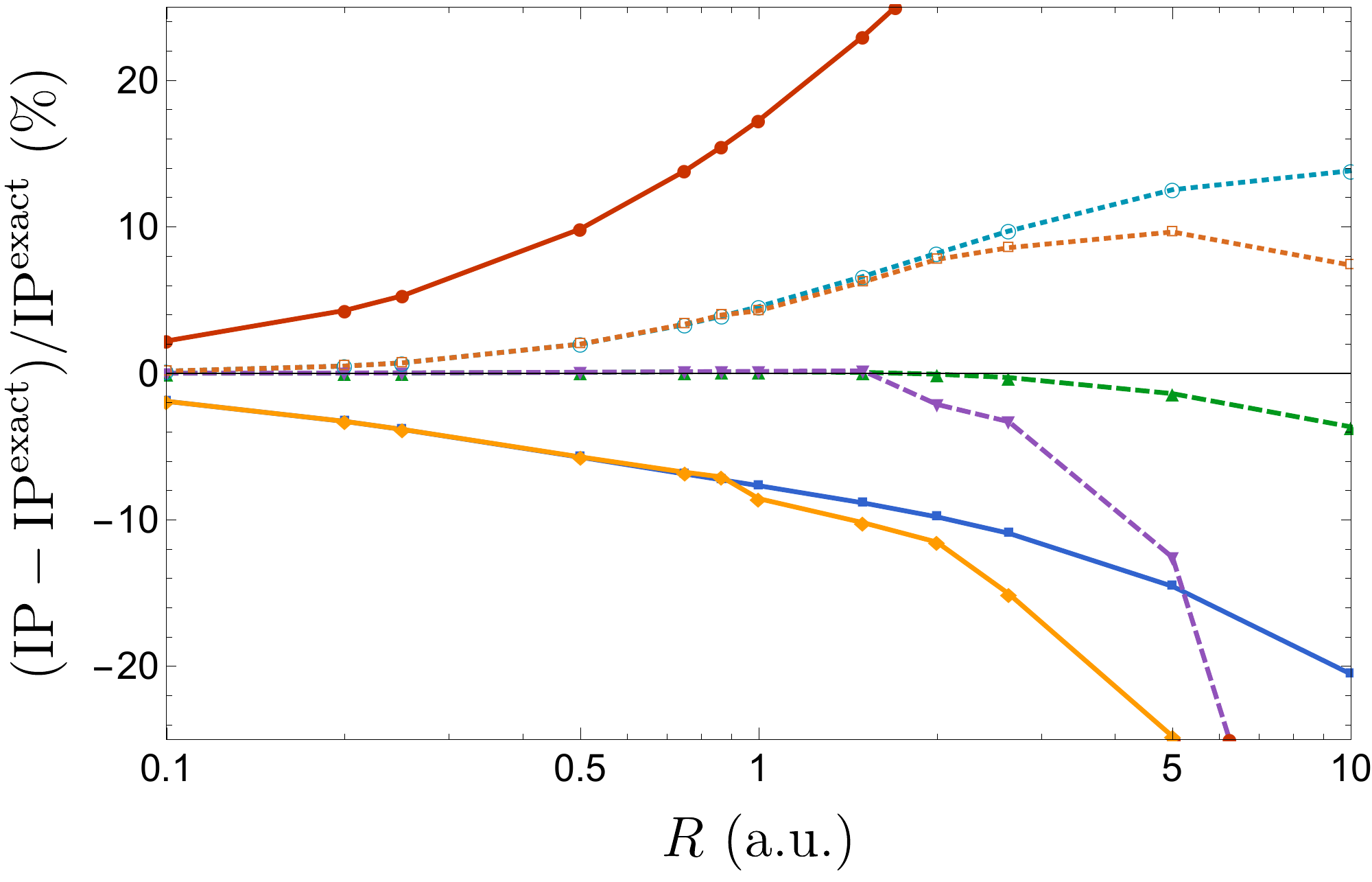}
	\includegraphics[width=\linewidth]{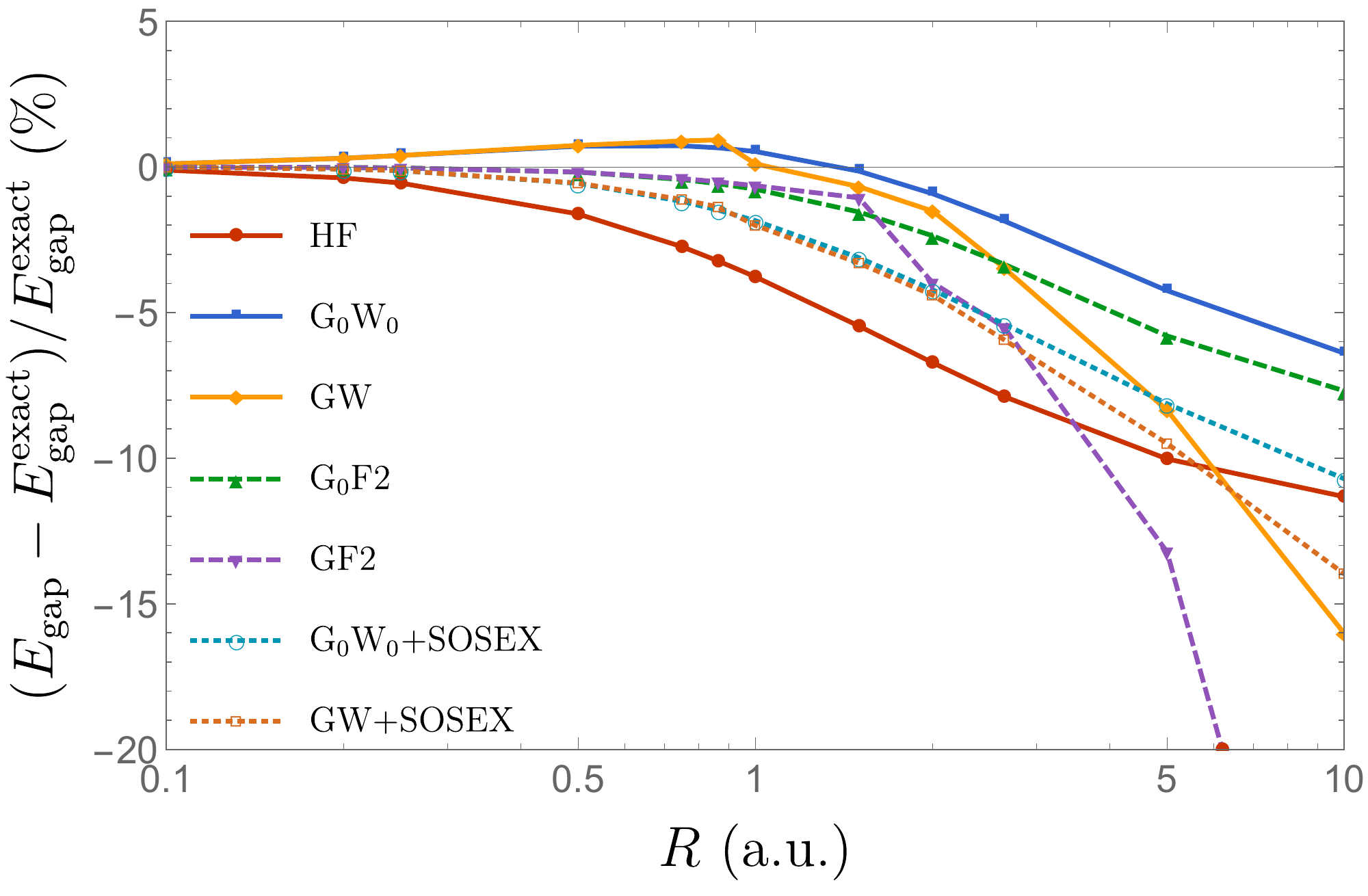}
	\includegraphics[width=\linewidth]{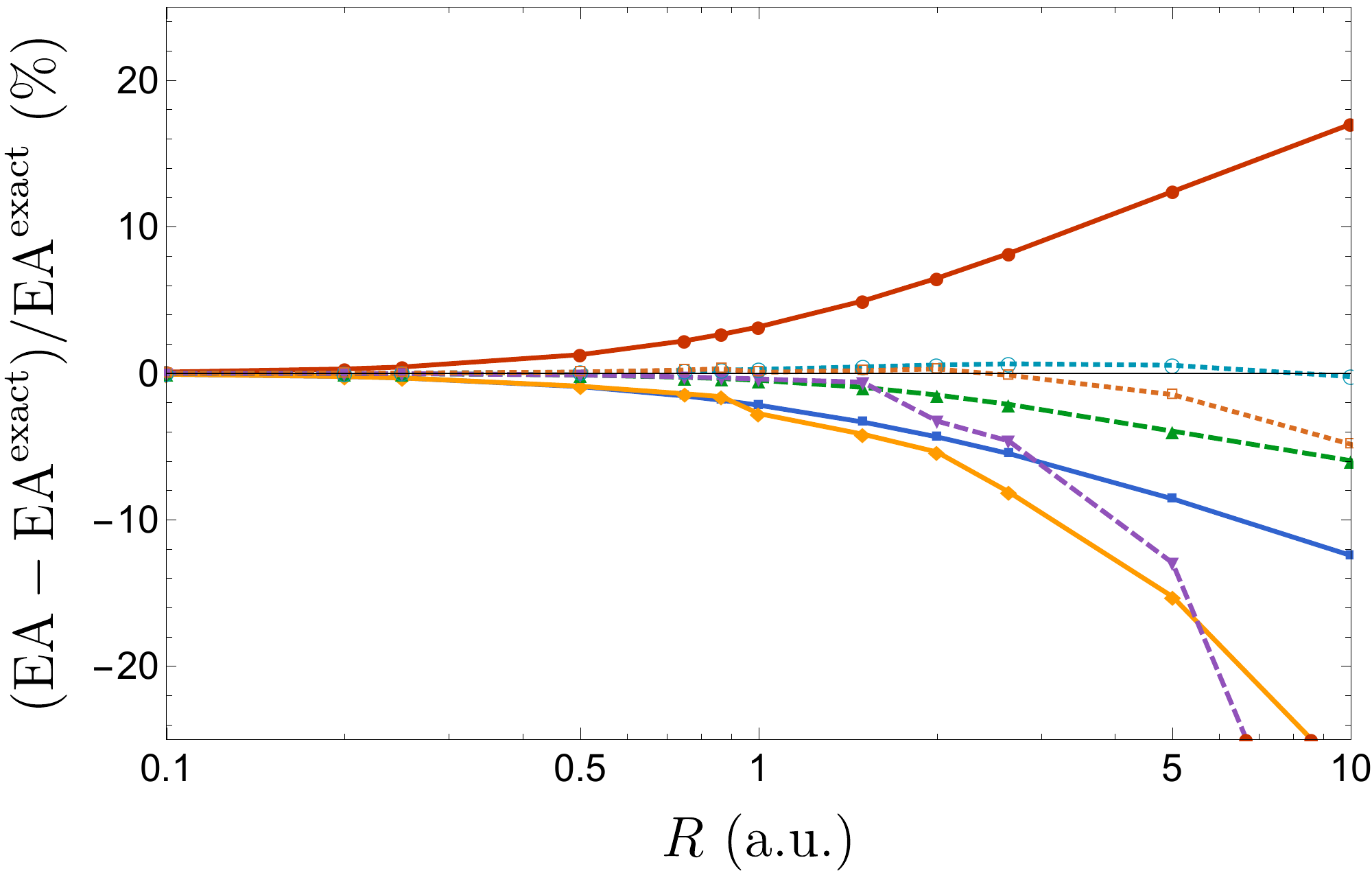}
	\caption{
	\label{fig:IPEAGapwrtR}
	Relative error (in \%) on IP (top), $\Egap$ (middle) and EA (bottom) as a function of $R$ for various schemes.
	See the Supporting Information for raw data.}
\end{figure}

\subsection{Ionization potential, electron affinity and energy gap}
The ionization potential (IP) and electron affinity (EA) are defined as \cite{SzaboBook}
\begin{align}
  \IP   & = -\eHOMO,
	&
  \EA   & = -\eLUMO,
\end{align}
where $\eHOMO$ and $\eLUMO$ are the HOMO ($\ell = 0$) and LUMO ($\ell = 1$) orbital energies, respectively, while the energy gap is
\begin{equation}
  \Egap = \eLUMO - \eHOMO = \IP - \EA.
\end{equation}
These results are shown in Fig.~\ref{fig:IPEAGapwrtR}, 
where we have reported the relative error (in \%) on $\IP$, $\EA$ and $\Egap$ as a function of $R$ for various methods from the weakly correlated regime ($R \ll 1$) to the strongly correlated regime ($R \gg 1$).
(The associated numerical results can be found in Tables I, II and III in the Supporting Information.)

The first striking observation is the quality of {\GOF} (dashed green curve in Fig.~\ref{fig:IPEAGapwrtR}), which yields accurate results up to $R \approx 2$, a regime in which the system can certainly be considered as weakly correlated.
Indeed, for two-electron systems, GF2 is known to be particularly accurate. \cite{Hernandez_1977, Casida_1989, Casida_1991}
However, within GF2 and GW, the effect of self-consistency is disappointing.
For example, the perturbative {\GOWO} version (solid blue curve) yields more accurate estimates than its self-consistent counterpart (solid yellow curve).
Similar observations can be made for GF2 (dashed lines) although the self-consistency starts deteriorating the results at larger $R$.

{\GOWO} and {\GW} are particularly bad at reproducing the ionization energies, even in the high-density (i.e.~small-$R$) limit.
The electron affinities are better reproduced, while $\Egap$ benefits from error cancelations (at least for small $R$).
This indicates that the poor performance of GW for the $\IP$ is mainly due to self-screening, i.e., the hole that is left behind after ionization is not just screened by the electron that remains but also by the electron that is removed.\cite{Romaniello_2009, Romaniello_2012, Tarantino_2017, Wetherell_2018}
This is clearly unphysical and happens only for the IP; for the EA the additional electron is correctly screened by both electrons.

Indeed, {\GWSOSEX} (dotted cyan line), which is mostly self-screening-free, does a good job for the $\IP$, although it still overestimates at large $R$.
The SOSEX correction also improves the $\EA$ but $\Egap$ is slightly less accurate for intermediate $R$ values.
In the case of {\GWSOSEX}, the effect of self-consistency is also disappointing and, except for the IP, deteriorates the results compared to the perturbative version.

Finally, we note that discontinuities appear around $R = 1$ and $2$ in the case of {\GW} and {\GF}, respectively.
We will discuss this in more detail in Sec.~\ref{sec:glitch})

\begin{figure*}
	\includegraphics[width=0.49\linewidth]{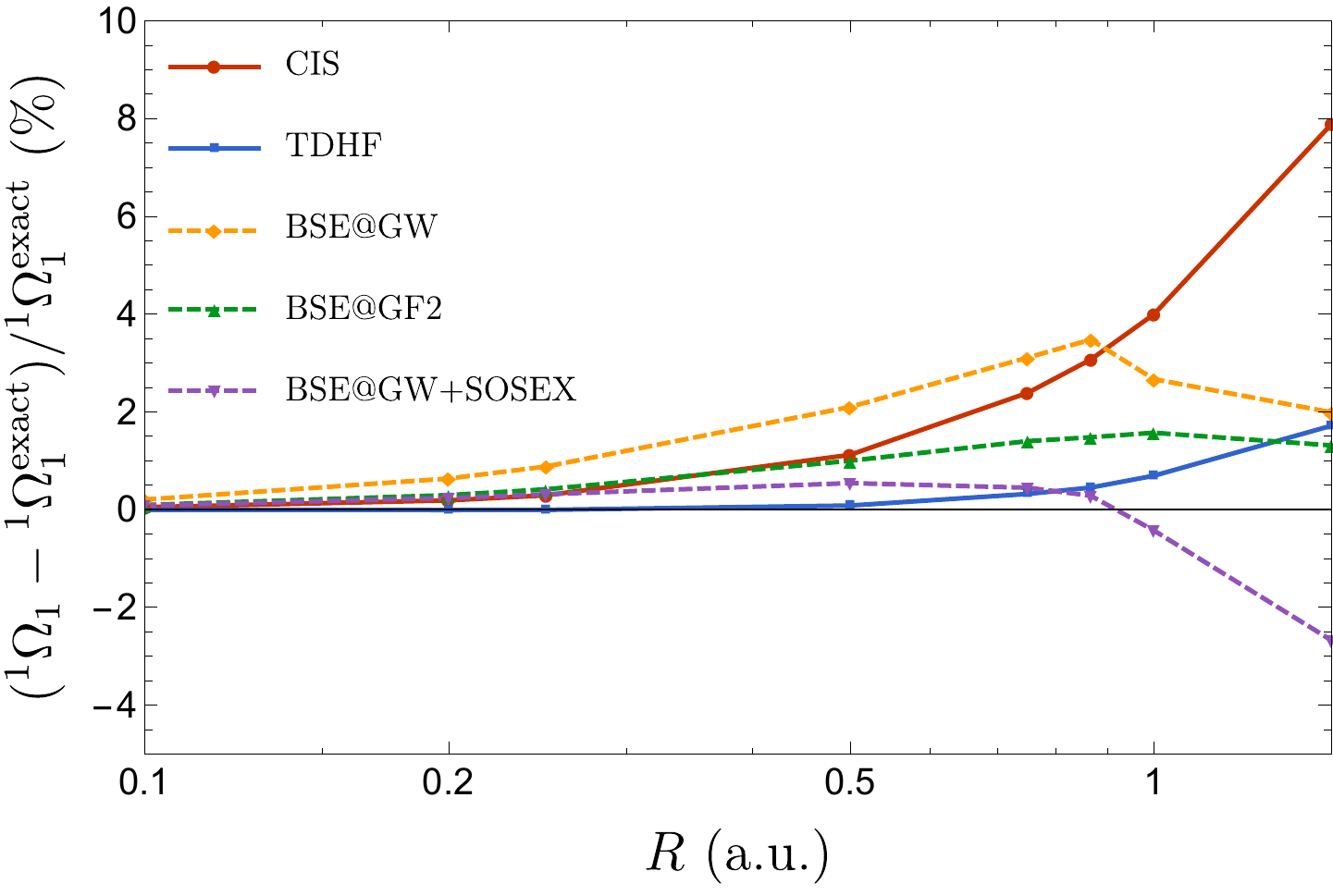}
	\includegraphics[width=0.49\linewidth]{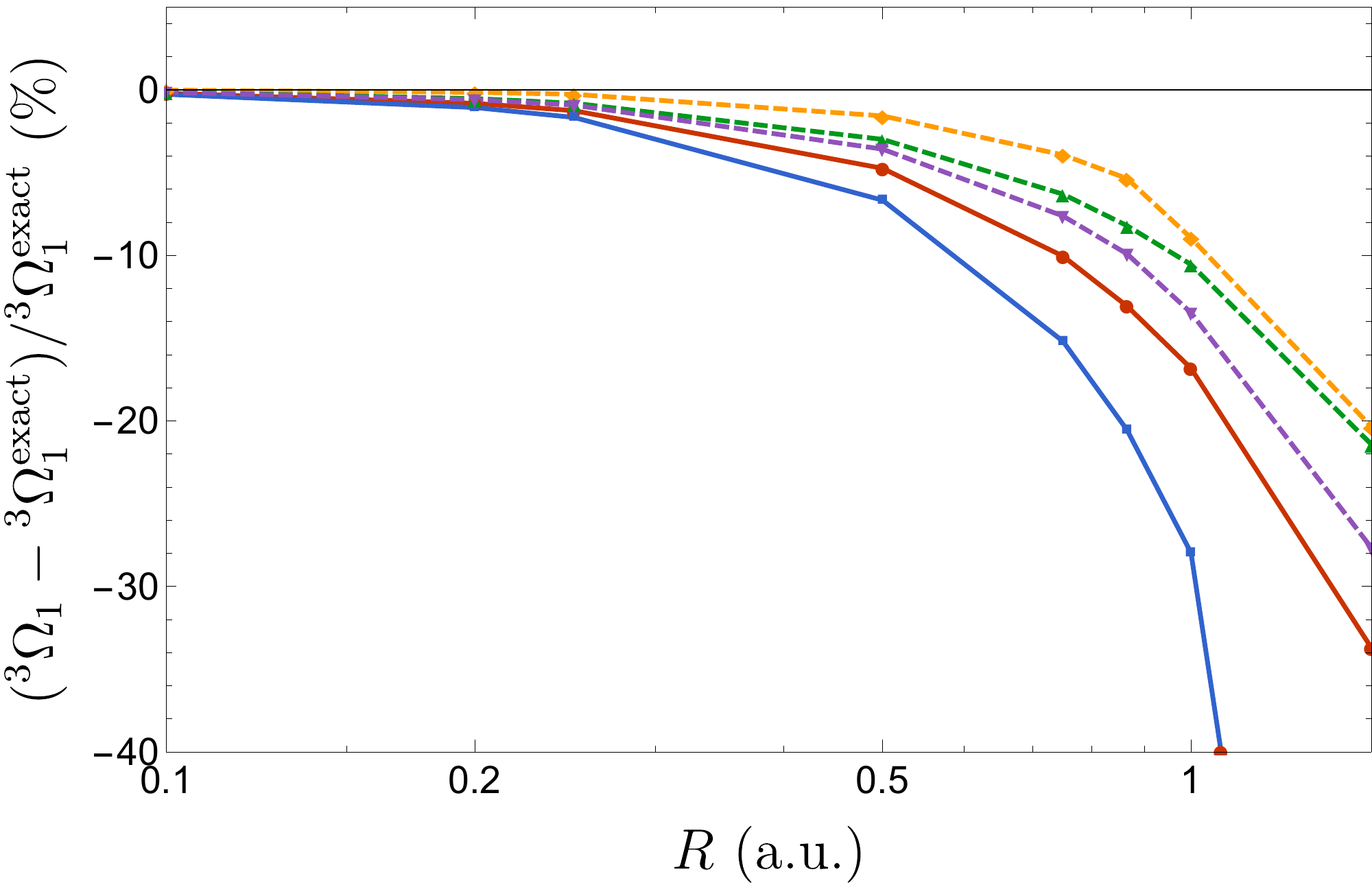}
	\caption{
	\label{fig:ExwrtR}
	Relative error (in \%) on the lowest singlet excitation $\Oms{1}$ (left) and the lowest triplet excitation $\Omt{1}$ (right) as a function of $R$ for various schemes. Note the different scales of the two graphs. See the Supporting Information for raw data.}
\end{figure*}

\subsection{
\label{sec:Ex}
Neutral excitations}
Figure \ref{fig:ExwrtR} shows the relative error (in \%) for the lowest singlet (left) and triplet (right) excitation energies (i.e.~$\ell = 1$) as a function of $R$ for various methods.
(The associated numerical results can be found in Tables IV and V in the Supporting Information.)
Because for $R > 3/2$ triplet instabilities appear due to the existence of a lower-energy (symmetry-broken) HF solution, \cite{Loos_2009a} we restrict our study to the high-density region $0 < R \le 3/2$. 

Because the GW eigenvalues are not significantly modified by the level of self-consistency in the high-density region (as one can see from Fig.~\ref{fig:IPEAGapwrtR}), we have chosen not to report the {\GOWO} curves which are very similar (yet not strictly identical) to the self-consistent GW ones depicted in Fig.~\ref{fig:ExwrtR}.
In other words, the BSE excitations do not depend strongly on the input GW eigenvalues.

Concerning the singlet manifold (left graph of Fig.~\ref{fig:ExwrtR}), TDHF is the most reliable method.
Although {BSE@\GW} appears as the least accurate method, it yields singlet excitation energies within a few percents of the FCI results.
BSE@{\GWSOSEX} shows significant improvement compared to the SOSEX-free methods.
Again, one can notice the discontinuity around $R = 1$ in the BSE@{\GW} curve (see below).

For the triplet manifold (right graph of Fig.~\ref{fig:ExwrtR}), BSE@{\GW} outperforms more conventional methods, such as CIS and TDHF as well as BSE@{\GWSOSEX} and BSE@{\GF}.
Note, however, that the magnitude of the errors for the triplet excitations are much larger than for the singlet ones.
This behavior is also observed in molecular systems, \cite{Jacquemin_2017} because of the inadequate singlet reference wave function used in most cases. \cite{Dreuw_2005} 

\begin{figure*}
	\includegraphics[width=\linewidth]{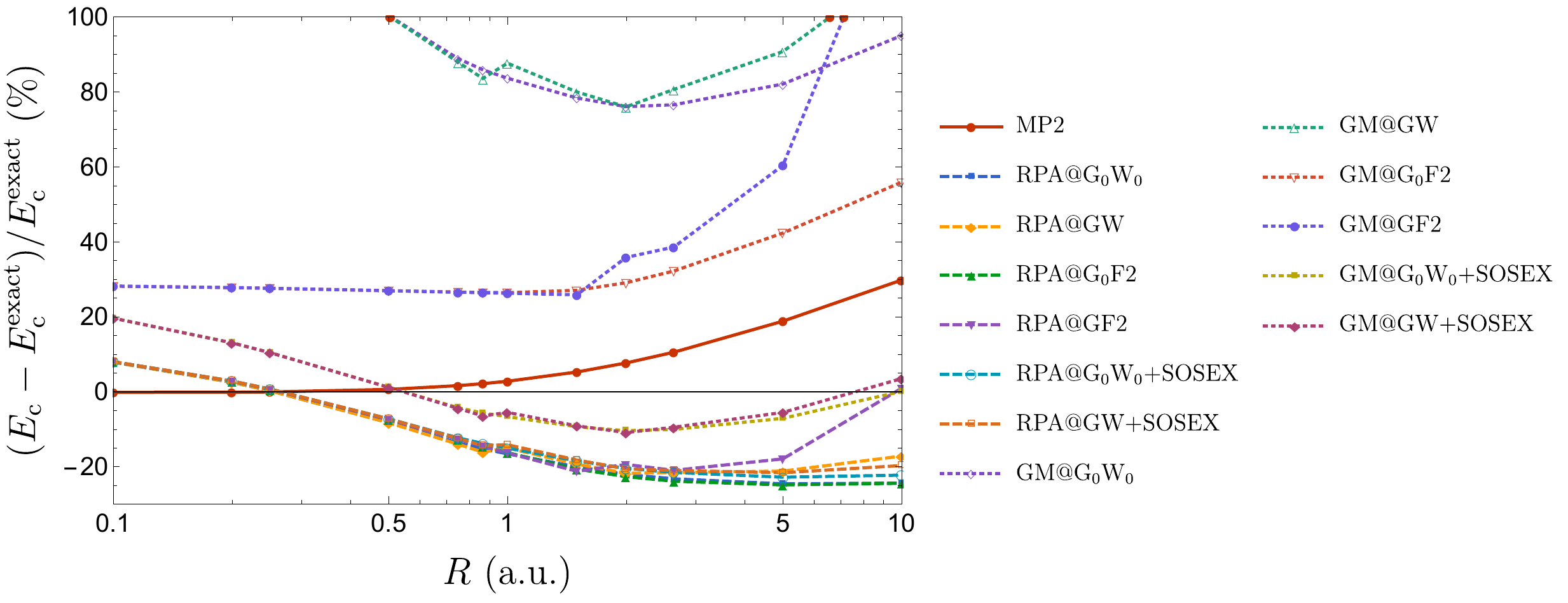}
	\caption{
	\label{fig:EcwrtR}
	Relative error (in \%) on the correlation energy $\Ec$ as a function of $R$ for various schemes.
	See the Supporting Information for raw data.}
\end{figure*}

\subsection{
\label{sec:Ec}
Correlation energy}
Concerning the correlation energy of spherium, the results are represented in Fig.~\ref{fig:EcwrtR}.
(The associated numerical results can be found in Table VI in the Supporting Information.)

MP2 (see Eq.~\eqref{eq:Ec-MP2}) provides a fairly consistent and reliable estimate of the correlation energy in the weakly correlated regime, although the relative error increases slightly when one gets to the strongly correlated regime where M{\o}ller-Plesset perturbation theory naturally breaks down. \cite{SzaboBook}

Firstly, let us mention that, in the weakly correlated regime, as expected, the correlation energies obtained with perturbative and self-consistent methods are very similar.
This has also been observed for atoms and molecules. \cite{Almbladh_1999, Stan_2006, Stan_2009, Dahlen_2004, Dahlen_2004a, Dahlen_2005, Dahlen_2005a, Dahlen_2006}
However, the situation is different in the strongly correlated regime and, for $R > 2$, the perturbative and its self-consistent variant start to deviate.

Because of its relation to the variational Klein functional, $\EcRPA$ is almost independent of the self-energy up to $R \approx 2$.
Interestingly, even in the large-$R$ regime, the RPA yields decent $\Ec$ estimates with a maximum error of ca.~20\%.

Unlike $\EcRPA$, the GM functional (known to be non-variational) is strongly dependent on the quality of $G$, and generally yields too negative correlation energies, an observation already made by several authors for the uniform electron gas \cite{Holm_1999, Holm_2000, Garcia-Gonzalez_2001}, solids \cite{Kutepov_2016, Kutepov_2017}, atoms and molecules. \cite{Stan_2006, Caruso_2013, Caruso_2013a, Caruso_2013b}
The self-screening in GW has a huge effect on $\EcGM$. 
For example, GM@GW is consistently wrong by about a factor two. 
When one improves the Green function, for instance by the introduction of second-order exchange, $\EcGM$ gets closer to the values obtained with $\EcRPA$.
In particular, GM@SOSEX, which removes most of the self-screening in GW, greatly improves the correlation energy, and even becomes more accurate than MP2 at large $R$.

For the correlation energies the influence of self-consistency is ambiguous.
While self-consistency improves the correlation energies in the case of $\EcRPA$, they deteriorate for $\EcGM$.

\begin{figure*}
	\includegraphics[width=0.7\linewidth]{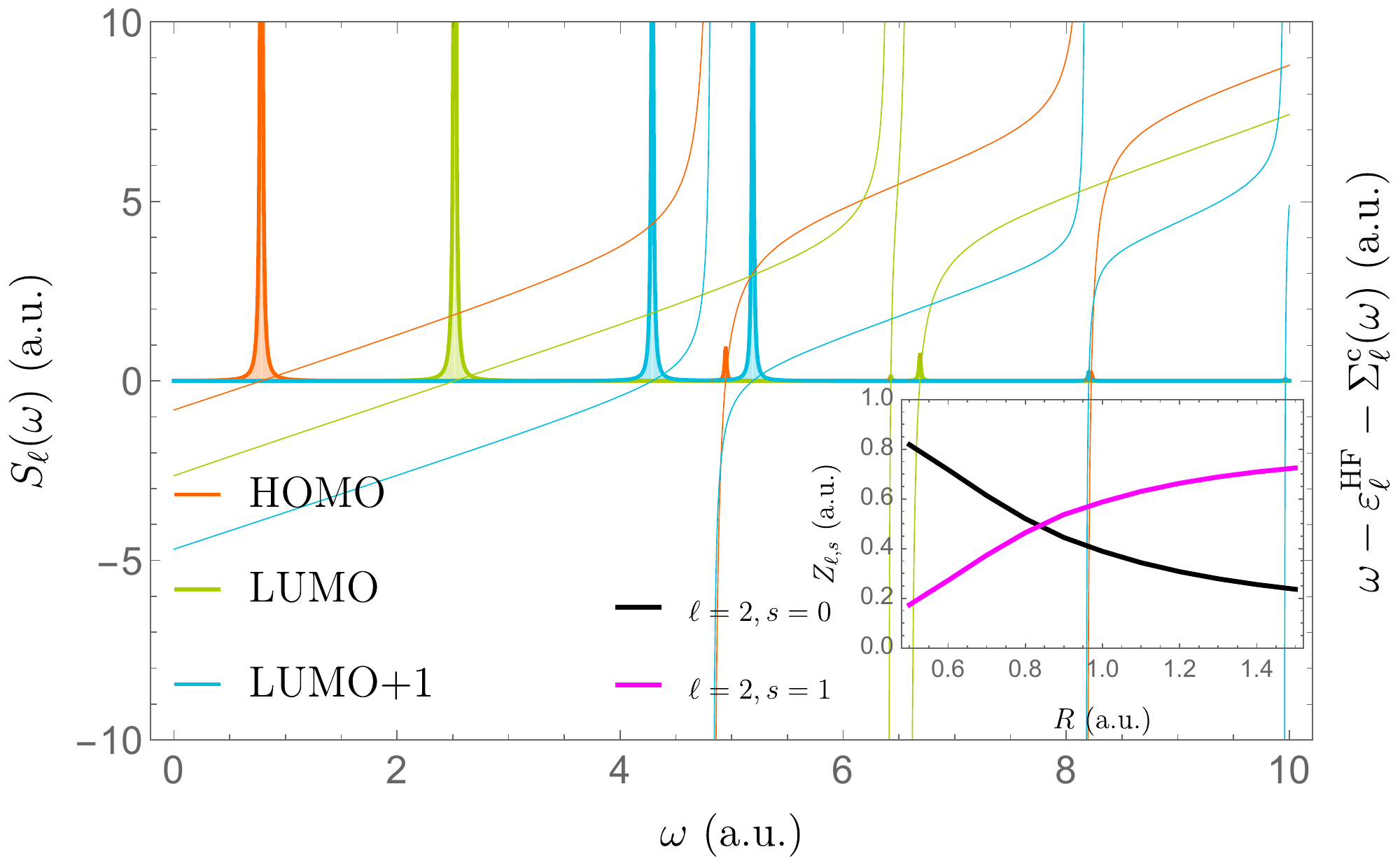}
	\caption{
	\label{fig:glitch}
	Self-consistent GW spectral function $\S{\ell}(\omega)$ as a function of $\omega$ (thick lines) for the HOMO (orange), LUMO (green) and LUMO+1 (cyan) orbitals at $R = 1$.
	The solutions of the quasiparticle equation are located at the intersection of the thin curves $\omega - \eHF{\ell} - \Re[\SigGW{\ell}(\omega)]$ and the horizontal axis.
	The inset graph reports the renormalization factor $\Z{\ell}(\e{\ell,s})$ of the LUMO+1 (i.e.~$\ell=2$) for the quasparticle peak ($s = 0$, black curve) and the first satellite ($s = 1$, magenta curve) as a function of $R$.
	}
\end{figure*}

\subsection{
\label{sec:glitch}
Binary system}
As mentioned several times earlier in this manuscript, there is an obvious discontinuity in Figs.~\ref{fig:IPEAGapwrtR}, \ref{fig:ExwrtR}, and \ref{fig:EcwrtR}) around $R \approx 0.9$.
Note that this ``glitch'' is only present in self-consistent calculations and is more pronounced in the SOSEX-free GW version.
Note also that its magnitude is small (yet numerically significant) and one would hardly notice it by looking at absolute energies.
From a technical point of view, the left and right sides of the discontinuity originate from two distinct solutions of the quasiparticle equation.
We note that this problem is different from the unphysical solutions discussed in Ref.~\onlinecite{Stan_2015}.

Our analysis has shown that this discontinuity is caused by the proximity of the quasiparticle peak of the LUMO+1 orbital (at $\e{2} \equiv \e{2,0} \approx 4.3$) and a singularity of $\SigGW{2}$ (at $\e{1} + \Om{1} \approx 4.8$).
This is illustrated in Fig.~\ref{fig:glitch} for a sphere of unit radius.
Due to the local symmetry of $\bSigGW$ at the vicinity of a singularity, it implies the existence of a satellite resonance (at $\e{2,1} \approx 5.2$) having a weight $\Z{2}(\e{2,1})$ of similar magnitude as the main quasiparticle peak $\Z{2}(\e{2,0})$. 
In that case, one cannot really talk about a quasiparticle peak and its satellite. 
It would be more appropriate to describe this peculiar situation as a binary system where both resonances have similar weights.
Figure \ref{fig:glitch} clearly shows the presence of two large-weight peaks for the LUMO+1 (thick cyan curve).
For these two peaks, we have reported, in the inset graph of Fig.~\ref{fig:glitch}, $\Z{\ell}(\e{\ell,s})$ as a function of $R$.
We see that, outside the range $1/2 \le R \le 3/2$, one resonance prevails over the other.
However, around $R = 1$, the quasiparticle peak ``morphs'' into a satellite and, vise versa, the satellite becomes the main quasiparticle.
Therefore, depending on the value of $R$, the self-consistent process selects one or the other peak depending on their relative weights.

To be best of our knowledge, this type of observation has never been reported in the literature.
Note that all this happens in the weakly correlated regime, where one should safely assume the validity of the quasiparticle picture.
We believe that such discontinuity would not exist within a fully self-consistent scheme where one takes into account the quasiparticle peak as well as its satellites at each iteration.
If confirmed, this would be a strong argument in favor of fully self-consistent schemes.
Finally, we note that these discontinuities are ubiquitous: they also appear for higher-energy orbitals and for larger radii.
We are currently analyzing the cause of such discontinuities in more details. 

\section{
\label{sec:conclusion}
Conclusion}
We have provided an exhaustive study of the performance of different commonly-used variants of Green function methods for the two-electron spherium model.
We found that, in general, self-consistency deteriorates the results with respect to those obtained within perturbative GW starting from Hartree-Fock orbital energies.
This is the case for many properties of interest, such as ionization potentials, electron affinities, and energy gaps.
Only for RPA correlation energies do we observe a small improvement when doing the calculations self-consistently.
We showed that the same is true for GF2, i.e, self-consistent GF2 results are, in general, worse than those obtained perturbatively.
We have also discussed the problem of self-screening in GW and showed that it can be partially cured by adding a second-order screened exchange (SOSEX) correction. 
We observe that this correction generally improve results. 
However, here again, self-consistency is disappointing.
Finally, we have evidenced that partially self-consistent GW can lead to artificial discontinuities in the self-energy.
We traced this problem back to the appearance of a satellite resonance with a weight similar to that of the quasiparticle.

\begin{acknowledgments}
The authors would like to thank Fabio Caruso, Fabien Bruneval, Denis Jacquemin and Xavier Blase for helpful discussions about GW methods.
\end{acknowledgments}

%

\end{document}